\documentclass[twocolumn]{jpsj3}

\usepackage{txfonts}
\usepackage{graphicx}
\usepackage{dcolumn}
\usepackage{bm}
\usepackage[usenames,dvipsnames]{xcolor}
\usepackage{ulem}
\usepackage{amsmath}
\usepackage{braket}
\usepackage{enumerate}
\usepackage{arydshln}
\usepackage{multirow}

\title{Multiple-$Q$ spin textures induced by spiral--staggered interference in one-dimensional itinerant magnets}
\author{Satoru Hayami$^1$, Kazuki Okigami$^2$, }
\inst{
  $^1$Graduate School of Science, Hokkaido University, Sapporo 060-0810, Japan\\
  $^2$Department of Applied Physics, The University of Tokyo, Tokyo 113-8656, Japan
 }
\abst{
We theoretically investigate multiple-$Q$ magnetic states emerging from the interference between finite-$Q$ spiral and staggered spin modulations in a one-dimensional itinerant electron system. 
The multiple-$Q$ spin textures are characterized by a superposition of symmetry-unrelated ordering wave vectors in the same direction with distinct periodicities rather than rotationally symmetry-related ones. 
Motivated by recent experimental observations of broken helix magnetic structures in EuIn$_2$As$_2$, we focus on the microscopic interaction conditions in stabilizing such multiple-$Q$ states.
We employ two effective spin models: One is the momentum-space-based model, and the other is the real-space-based model, both of which include bilinear and biquadratic easy-plane anisotropic interactions. 
By analyzing their ground state via simulated annealing, we find that a superposition of a spiral and a staggered modulation yields a robust double-$Q$ magnetic structure. 
Moreover, we demonstrate that the obtained double-$Q$ spin configuration exhibits an antisymmetric spin-split band structure even without the relativistic Dzyaloshinskii-Moriya interaction, and further reveals asymmetric band modulations when the magnetic field is applied along the out-of-plane direction. 
Our results provide a theoretical framework for understanding unconventional multiple-$Q$ magnetic textures in one-dimensional systems.
}

\begin{document}
\maketitle

\section{Introduction}
\label{sec:introduction}

Multiple-$Q$ magnetic states, which often lead to nontrivial spin textures composed of the superposition of spin modulations with distinct ordering vectors~\cite{Bak_PhysRevLett.40.800, McEwen_PhysRevB.34.1781, zochowski1986thermal, forgan1990magnetic, Longfield_PhysRevB.66.054417, Bernhoeft_PhysRevB.69.174415, stewart2004phase, Watson_PhysRevB.53.726, Harris_PhysRevB.74.134411}, have become a central topic in condensed matter physics~\cite{nagaosa2013topological}. 
These states often host unconventional physical phenomena as a consequence of vector and scalar spin chiralities in noncollinear and noncoplanar spin textures, such as topological Hall effects~\cite{Ohgushi_PhysRevB.62.R6065,taguchi2001spin,tatara2002chirality, Neubauer_PhysRevLett.102.186602, Hamamoto_PhysRevB.92.115417,nakazawa2018topological, tai2022distinguishing} and magnetoelectric responses~\cite{Katsura_PhysRevLett.95.057205, Mostovoy_PhysRevLett.96.067601, SergienkoPhysRevB.73.094434, Harris_PhysRevB.73.184433,tokura2014multiferroics,cardias2020first, Hayami_PhysRevB.105.104428, Bhowal_PhysRevLett.128.227204}. 
Multiple-$Q$ states have been mainly studied in higher-dimensional systems where the constituent ordering wave vectors are related by rotational symmetry of the crystal structure, such as $C_6$ or $C_4$ rotations, resulting in symmetric superpositions across reciprocal space~\cite{Bak_PhysRevLett.40.800, Shapiro_PhysRevLett.43.1748,bak1980theory, Forgan_PhysRevLett.62.470, hayami2024stabilization}.
For example, a superposition of sixfold ordering wave vectors gives rise to a triple-$Q$ triangular skyrmion crystal~\cite{Muhlbauer_2009skyrmion,yu2010real, rossler2006spontaneous, Binz_PhysRevLett.96.207202,  Binz_PhysRevB.74.214408, Yi_PhysRevB.80.054416}, whereas that of fourfold ordering wave vectors gives rise to a double-$Q$ square skyrmion crystal~\cite{Yi_PhysRevB.80.054416, khanh2020nanometric, singh2023transition}. 

Meanwhile, recent developments have highlighted the existence of multiple-$Q$ states beyond symmetry-related conditions. 
One remarkable example is the broken helix spin structure observed in the layered magnetic compound EuIn$_2$As$_2$, where a helical component coexists with a commensurate staggered spin modulation~\cite{riberolles2021magnetic, soh2023understanding, Donoway_PhysRevX.14.031013, takeda2024incommensurate, Gen_PhysRevB.111.L081109}. 
Unlike conventional multiple-$Q$ states arising from symmetry-related ordering vectors, this broken helix state involves the interference of two wave vectors that are not connected by symmetry, but rather by a uniaxial one-dimensional relationship consisting of an incommensurate wave vector and a staggered one. 
This observation raises intriguing questions about the stability condition of complex multiple-$Q$ states based on the microscopic modeling.

Motivated by such findings, in this study, we theoretically investigate a one-dimensional type of multiple-$Q$ state that emerges from the interference between spiral and staggered magnetic orders, whose ordering wave vectors are located in the same spatial direction, in an itinerant electron system. 
By performing simulated annealing for an effective spin model incorporating momentum-resolved bilinear and biquadratic easy-plane anisotropic interactions in the centrosymmetric one-dimensional spin chain, we investigate the ground-state phase diagram in a wide range of model parameters. 
As a result, we show the two important conditions to invoke the double-$Q$ instability irrespective of the magnitude of ordering wave vectors: One is the large biquadratic interaction, and the other is the competition of the bilinear interaction at finite-$Q$ and staggered wave-vector channels. 
We confirm such a tendency by analyzing another model with real-space bilinear and biquadratic interactions. 
We also demonstrate that the double-$Q$ state can exhibit the antisymmetric spin splitting at zero magnetic field and asymmetric band modulation in a finite magnetic field as a consequence of its spiral spin texture. 
Our findings shed new light on the landscape of magnetic orders in one-dimensional magnetic systems and provide a theoretical framework for understanding nontrivial spin configurations such as the broken helix state observed in EuIn$_2$As$_2$. 
Furthermore, the present work suggests that further complex multiple-$Q$ states, such as a superposition of the broken helix states along different spatial directions, may be possible in two- and three-dimensional systems, even in the absence of the relativistic Dzyaloshinskii-Moriya interaction~\cite{dzyaloshinsky1958thermodynamic,moriya1960anisotropic}.

The remainder of the present paper is organized as follows. 
In Sec.~\ref{sec: Model and method}, we introduce an effective spin model with momentum-resolved exchange interactions. 
We also outline the numerical method based on simulated annealing. 
In Sec.~\ref{sec: Multiple-$Q$ instability}, we construct the ground-state phase diagram in the momentum-resolved spin model and show the parameter region where the double-$Q$ state emerges. 
In Sec.~\ref{sec: Real-space simulations}, we introduce the real-space spin model and show its ground state. 
Then, we discuss electronic band structures under the double-$Q$ state as well as that under the single-$Q$ spiral state in Sec.~\ref{sec: Electronic band structure}. 
Finally, Sec.~\ref{sec:summary} summarizes the present results.

\section{Momentum-resolved spin model and method}
\label{sec: Model and method}

We consider the following effective momentum-resolved spin model in the one-dimensional chain system with the spatial inversion symmetry (the lattice constant is set as unity), which is given by~\cite{Hayami_PhysRevB.95.224424, Yambe_PhysRevB.106.174437} 
\begin{align}
\label{eq: Ham}
\mathcal{H}^{\rm eff}=& - J_{\rm sp} \sum_{\nu=x,y} S^\nu_{Q_{\rm sp}}  S^\nu_{-Q_{\rm sp}} - J_{\rm sta} \sum_{\nu=x,y} S^\nu_{Q_{\rm sta}}  S^\nu_{-Q_{\rm sta}} 
 \nonumber \\
&+ \frac{K_{\rm sp}}{N} \left(\sum_{\nu=x,y} S^\nu_{Q_{\rm sp}}  S^\nu_{-Q_{\rm sp}}\right)^2+ \frac{K_{\rm sta}}{N} \left(\sum_{\nu=x,y} S^\nu_{Q_{\rm sta}}  S^\nu_{-Q_{\rm sta}}\right)^2, 
\end{align}
where $S^\nu_{q}$ represents the spin moment with the component $\nu=x,y$ and the wave vector $q$, which is related to real-space spin $\bm{S}_i = (S^x_i, S^y_i)$ with $|\bm{S}_i|=1$ via the Fourier transformation; we adopt the $xy$-type Heisenberg spin model by suppsoing the large easy-plane anisotropic interaction bearing in mind the situation in EuIn$_2$As$_2$~\cite{Gen_PhysRevB.111.L081109}. 
While the broken-helix state in EuIn$_2$As$_2$ motivates this work, our analysis focuses on elucidating the general conditions required to stabilize the double-$Q$ broken helix state, without addressing material-specific microscopic details.

The first term in Eq.~(\ref{eq: Ham}) stands for the bilinear exchange interaction at a finite-$Q$ wave vector. 
We only choose one of the finite-$Q$ interaction channels with $0 < Q_{\rm sp} < \pi$ for simplicity. 
This interaction tends to lead to the instability toward the single-$Q$ spiral state with $Q_{\rm sp}$. 
The second term stands for the bilinear exchange interaction at $Q_{\rm sta}=\pi$, which favors the staggered alignment of spins. 
The third and fourth terms denote the positive biquadratic interactions at $Q_{\rm sp}$ and $Q_{\rm sta}$, respectively. 
This biquadratic interaction tends to induce topologically nontrivial multiple-$Q$ instabilities with the symmetry-related ordering wave vectors even in the absence of an external magnetic field, such as the high-topological-number skyrmion crystal~\cite{Hayami_PhysRevB.95.224424, Hayami_PhysRevB.99.094420, hayami2020multiple, hayami2021phase, hayami2022multiple}, meron-antimeron crystal~\cite{Hayami_PhysRevB.104.094425}, and hedgehog crystal~\cite{Okumura_PhysRevB.101.144416, Okumura_doi:10.7566/JPSJ.91.093702} in two- and three-dimensional systems; $N$ stands for the total number of spins. 

This effective spin model is derived from the Kondo lattice model by tracing out the charge degree of freedom of itinerant electrons, where $J_{\rm sp}$ and $J_{\rm sta}$ ($K_{\rm sp}$ and $K_{\rm sta}$) are derived as a consequence of the lowest second-order (second-lowest fourth-order) perturbations~\cite{Ruderman, Kasuya, Yosida1957, Akagi_PhysRevLett.108.096401, Hayami_PhysRevB.90.060402}, which is used to reproduce the experimental phase diagram hosting complicated multiple-$Q$ states, such as the triple-$Q$ vortex in Y$_3$Co$_8$Sn$_4$~\cite{takagi2018multiple} and the field-sensitive skyrmion crystal in EuPtSi~\cite{kakihana2018giant,kaneko2019unique,tabata2019magnetic, kakihana2019unique, hayami2021field}. 
The lowest second-order interaction is often referred to as the Ruderman-Kittel-Kasuya-Yosida (RKKY) interaction~\cite{Ruderman, Kasuya, Yosida1957}.

It should, however, be noted that Eq.~(\ref{eq: Ham}) is not intended as a microscopic Hamiltonian derived uniquely from a specific electronic model. 
Rather, it is introduced as a phenomenological minimal model that captures the competition between two distinct ordering tendencies at $Q_{\rm sp}$ and $Q_{\rm sta}$. 
In particular, the choice of these two ordering wave numbers does not assume the existence of two Fermi wave numbers; instead, it provides a minimal parametrization to describe the coexistence of spiral and staggered components observed in frustrated spin systems.
Similarly, the positive biquadratic terms $K_{\rm sp}$ and $K_{\rm sta}$ are consistent with higher-order RKKY-type interactions but are employed phenomenologically to stabilize noncollinear and multiple-$Q$ configurations, including the broken helix state.

In the following, we systematically change these phenomenological parameters in order to investigate the ground state of the model in Eq.~(\ref{eq: Ham}) in a wide range of parameters. 
We neglect other magnetic anisotropic interactions like the bond-dependent anisotropic interactions and dipole-dipole interactions, which sometimes lead to multiple-$Q$ instabilities~\cite{Hayami_PhysRevB.103.054422, amoroso2020spontaneous, yambe2021skyrmion, Wang_PhysRevB.103.104408, Utesov_PhysRevB.103.064414, amoroso2021tuning, Utesov_PhysRevB.105.054435}.

To investigate multiple-$Q$ instabilities in the one-dimensional spin-chain model, we implicitly consider the chain to be embedded in a three-dimensional system with weak ferromagnetic interchain coupling.
We construct low-temperature magnetic phase diagrams while varying $J_{\rm sta}$ and $Q_{\rm sp}$ for several $K \equiv K_{\rm sp}$. 
We set $J_{\rm sp}=1$ as the energy unit of the model and $K_{\rm sta}=K(J_{\rm sta}/J_{\rm sp})^2$. 
For the system size, we set $N=200$ under the periodic boundary conditions. 
As the numerical method, we adopt simulated annealing in the following manner. 
The simulations are initialized from random spin configurations at high temperatures $T_0=$1--5, and the temperature is gradually reduced according to the annealing schedule $T_{n+1} = 0.999999 T_n$ until a final temperature of $T = 0.001$ is reached. 
At each temperature step, local spin updates are carried out sequentially using the standard Metropolis algorithm. 
After reaching the final temperature, additional Monte Carlo sweeps on the order of $10^5$--$10^6$ are performed to obtain physical observables. 
For each set of model parameters, independent annealing runs are conducted. 

The obtained spin configurations are classified by the $q$ component of momentum-resolved spins, which is given by 
\begin{align}
m_{q}&=\frac{1}{N}\sqrt{\sum_{\nu=x,y}\sum_{ij}S_i^\nu S_j^\nu e^{i q  (x_i -x_j)}}, 
\end{align}
where $x_i$ is the position vector at site $i$. 
Owing to the constraint in terms of the local spin length $|\bm{S}_i|=1$, the sum rule $\sum_q (m_q)^2 = 1$ is satisfied.  

\vspace{1cm}

\section{Multiple-$Q$ instability}
\label{sec: Multiple-$Q$ instability}

\begin{figure}[tb!]
\begin{center}
\includegraphics[width=1.0\hsize]{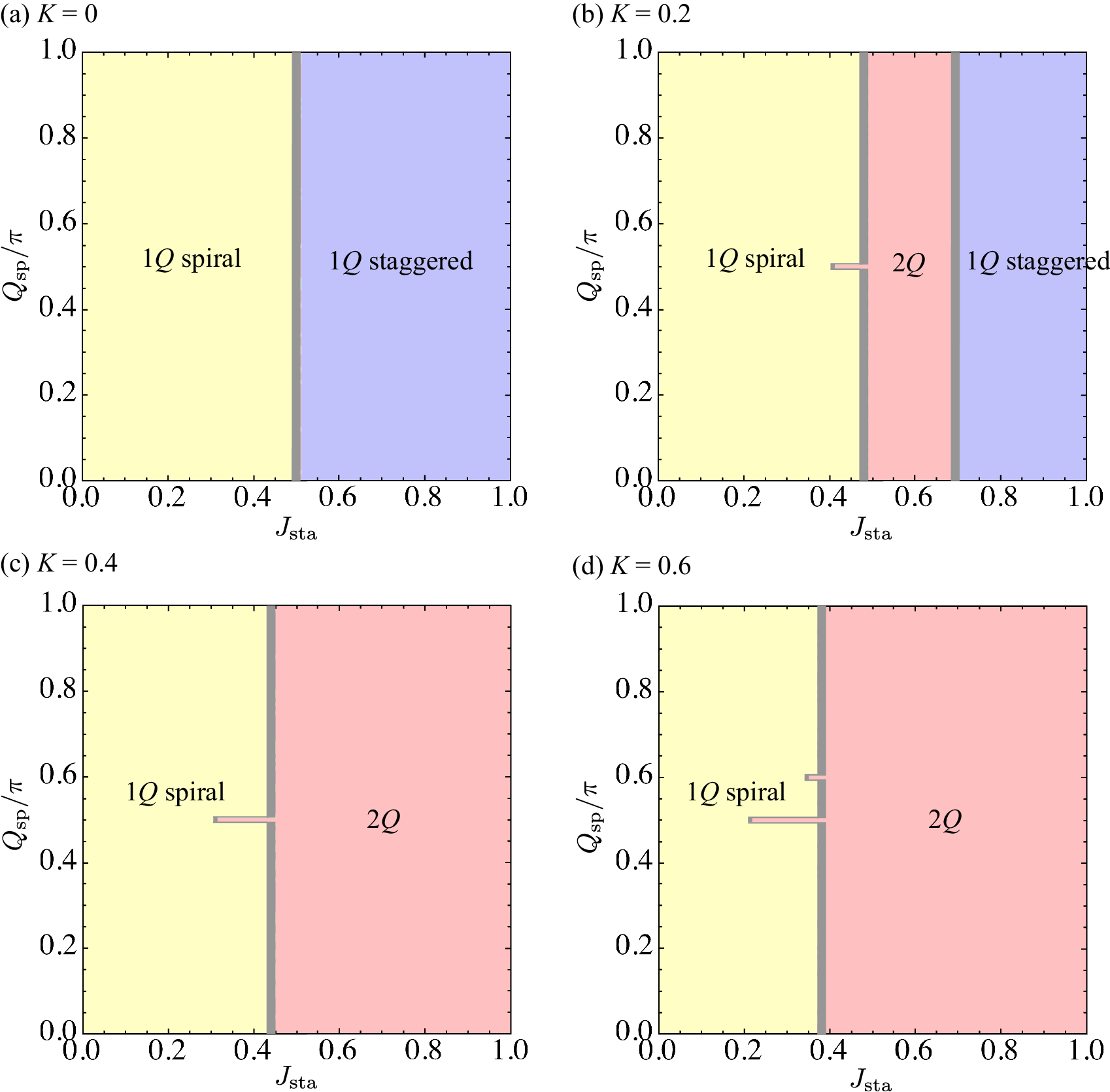} 
\caption{
\label{fig: PD} 
(Color online) Ground-state phase diagram in the plane of $J_{\rm sta}$ and $Q_{\rm sp}$ at (a) $K=0$, (b) $K=0.2$, (c) $K=0.4$, and (d) $K=0.6$. 
The thick gray lines denote the phase boundaries. 
1$Q$ and 2$Q$ stand for the single-$Q$ and double-$Q$ states, respectively. 
}
\end{center}
\end{figure}

\begin{figure}[tb!]
\begin{center}
\includegraphics[width=1.0\hsize]{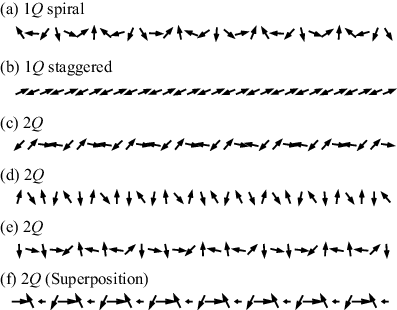} 
\caption{
\label{fig: spin} 
Real-space spin configurations of (a) the single-$Q$ spiral state at $K=0$, $J_{\rm sta}=0.4$, and $Q_{\rm sp}=0.3\pi$, (b) the single-$Q$ staggered state at $K=0$, $J_{\rm sta}=0.6$, and $Q_{\rm sp}=0.3\pi$, (c) the double-$Q$ state at $K=0.2$, $J_{\rm sta}=0.44$, and $Q_{\rm sp}=0.5\pi$, (d) the double-$Q$ state at $K=0.2$, $J_{\rm sta}=0.6$, and $Q_{\rm sp}=0.3\pi$, and (e) the double-$Q$ state at $K=0.6$, $J_{\rm sta}=0.4$, and $Q_{\rm sp}=0.2\pi$, which are obtained by simulated annealing. 
The spin configuraion in (f) stands for the simply superposed spin configuration as $\bm{S}_i= (\cos Q_{\rm sp} x_i + 0.5 \cos Q_{\rm sta} x_i, \sin Q_{\rm sp} x_i)$ without normalization ($|\bm{S}_i| \neq 1$) at $Q_{\rm sp}=0.5\pi$ for reference. 
}
\end{center}
\end{figure}

\begin{figure}[tb!]
\begin{center}
\includegraphics[width=1.0\hsize]{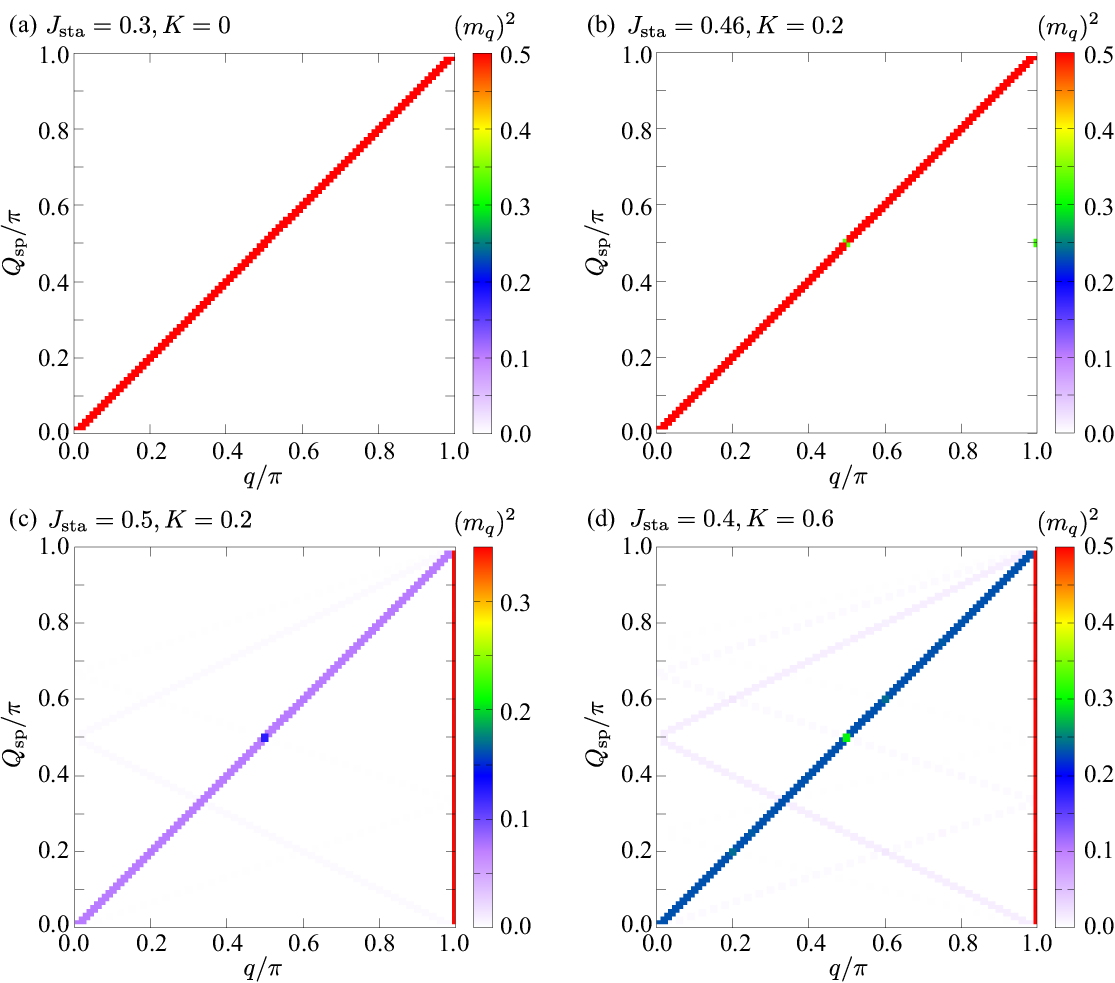} 
\caption{
\label{fig: Sq} 
(Color online) $Q_{\rm sp}$ dependence of the intensity plot of $(m_q)^2$ at (a) $J_{\rm sta}=0.3$ and $K=0$, (b) $J_{\rm sta}=0.46$ and $K=0.2$, (c) $J_{\rm sta}=0.5$ and $K=0.2$, and (d) $J_{\rm sta}=0.4$ and $K=0.6$. 
}
\end{center}
\end{figure}

To systematically examine the tendency of double-$Q$ instability, we consider 99 possible choices of $Q_{\rm sp}$ satisfying $n\pi/100 $ ($n=1$--$99$) for the system size $N=200$. 
Figure~\ref{fig: PD}(a) shows the low-temperature phase diagram against $J_{\rm sta}$ and $Q_{\rm sp}$ at $K=0$. 
The phase diagram consists of two phases separated by $J_{\rm sta}=0.5$; the single-$Q$ spiral state appears for $J_{\rm sta}<0.5$, while the single-$Q$ staggered state appears for $J_{\rm sta}>0.5$. 
The real-space spin configurations for the single-$Q$ spiral and single-$Q$ staggered states are presented in Figs.~\ref{fig: spin}(a) and \ref{fig: spin}(b), respectively. 
Thus, there is no double-$Q$ instability for $K=0$, which is understood from the fact that the bilinear exchange energy is maximized by the single-$Q$ spiral (or staggered) state according to the Luttinger-Tisza analysis~\cite{Luttinger_PhysRev.70.954}. 
Such a feature holds irrespective of $Q_{\rm sp}$, as shown in Fig.~\ref{fig: Sq}(a), where only the single-$Q$ peak structure is found. 

When the biquadratic interaction is introduced, the double-$Q$ state appears in the region sandwiched by the single-$Q$ spiral and single-$Q$ staggered states, as shown in the case of $K=0.2$ in Fig.~\ref{fig: PD}(b). 
Except for $Q_{\rm sp}=0.5\pi$, the critical value of $J_{\rm sta}$ from the single-$Q$ spiral state to the double-$Q$ state is the same. 
On the other hand, its critical value of $J_{\rm sta}$ for $Q_{\rm sp}=0.5\pi$ is smaller, which indicates that the double-$Q$ state with $Q_{\rm sp}=0.5\pi$ is a special case. 
This is because the higher-harmonic wave vectors constructed by a superposition of $Q_{\rm sp}=0.5\pi$ and $Q_{\rm sta}=\pi$ correspond to themselves, which means no energy loss by exchange energy in constructing a double-$Q$ superposition. 
Accordingly, the double-$Q$ state with $Q_{\rm sp}=0.5\pi$ tends to be realized compared to that with other $Q_{\rm sp}$, which is demonstrated in the case of $J_{\rm sta}=0.46$ and $K=0.2$ in Fig.~\ref{fig: Sq}(b).  
The real-space spin configuration for the double-$Q$ state with $Q_{\rm sp}=0.5\pi$ at $J_{\rm sta}=0.44$ is shown in Fig.~\ref{fig: spin}(c). 
For reference, we also show the real-space plot of the spin configuration obtained from the ansatz $\bm{S}_i= (\cos Q_{\rm sp} x_i + 0.5 \cos Q_{\rm sta} x_i, \sin Q_{\rm sp} x_i)$ in Fig.~\ref{fig: spin}(f).

For general $Q_{\rm sp} \neq 0.5\pi$, the higher harmonics consisting of $Q_{\rm sp}$ and $Q_{\rm sta}$ appear in the double-$Q$ spin structure stabilized for $0.48 \lesssim J_{\rm sta} \lesssim 0.7$ in Fig.~\ref{fig: PD}(b). 
The higher-harmonic wave vector emerges at $Q_{\rm sta}\pm 2Q_{\rm sp}$, as shown in Fig.~\ref{fig: Sq}(c), which modulates the real-space spin configuration from the purely single-$Q$ spiral and staggered states shown in Fig.~\ref{fig: spin}(d). 
The contributions from $Q_{\rm sta}\pm 2Q_{\rm sp}$ imply the spin ansatz of the double-$Q$ state as follows: 
\begin{align}
\bm{S}_i=\left\{\sqrt{1-b^2+b^2 \cos^2 (Q_{\rm sp}x_i)}\cos (Q_{\rm sta}x_i), b \sin (Q_{\rm sp}x_i)  
\right\}, 
\end{align}
where $0< b < 1$ is the parameter to determine the degree of the double-$Q$ structure.

We discuss the origin and the role of this ansatz. 
The analytic form above is constructed by referring to the Fourier components obtained from our numerical simulations, which clearly show higher harmonics at $Q_{\rm sta} \pm 2Q_{\rm sp}$. 
The functional form also follows the superposition scheme used for double-$Q$ states in Ref.~\citen{Ozawa_doi:10.7566/JPSJ.85.103703}, where two ordering wave vectors are symmetry related. 
The resulting structure represents the generic double-$Q$ modulation that naturally emerges when the positive biquadratic interaction becomes finite, regardless of the symmetry relation between the two ordering wave vectors, because such a superposition maximizes the energy gain associated with higher-order interactions~\cite{Hayami_PhysRevB.95.224424}. 
Importantly, the ansatz is not introduced merely for visualization purposes; it serves as an analytic representation that captures the essential features of the obtained double-$Q$ state. 
We also confirm that this ansatz reproduces the numerically obtained spin textures for a wide range of ordering vectors $(Q_{\rm sp}, Q_{\rm sta})$, demonstrating its robustness beyond specific parameter sets.

By further increasing $K$, the region for the double-$Q$ state is extended, and accordingly, the region for the single-$Q$ staggered state vanishes, as shown in the case of $K=0.4$ in Fig.~\ref{fig: PD}(c).  
The region for the double-$Q$ state is further enlarged for $K=0.6$, as shown in Fig.~\ref{fig: PD}(d). 
As $K$ increases, the double-$Q$ spin configuration is modulated as follows; the constituent $Q_{\rm sp}$ component changes from a spiral configuration to a sinusoidal one, and the oscillation direction of the sinusoidal component tends to become orthogonal to the direction of the staggered component. 
Such a change from the spiral to sinusoidal oscillation leads to the enhancement of the threefold periodic oscillation in terms of $Q_{\rm sp}$, i.e., $q = 3Q_{\rm sp}$, as shown in Fig.~\ref{fig: Sq}(d). 
A typical real-space spin configuration for large $K$ is shown in Fig.~\ref{fig: spin}(e). 

\section{Real-space simulations}
\label{sec: Real-space simulations}

\begin{figure}[tb!]
\begin{center}
\includegraphics[width=0.8\hsize]{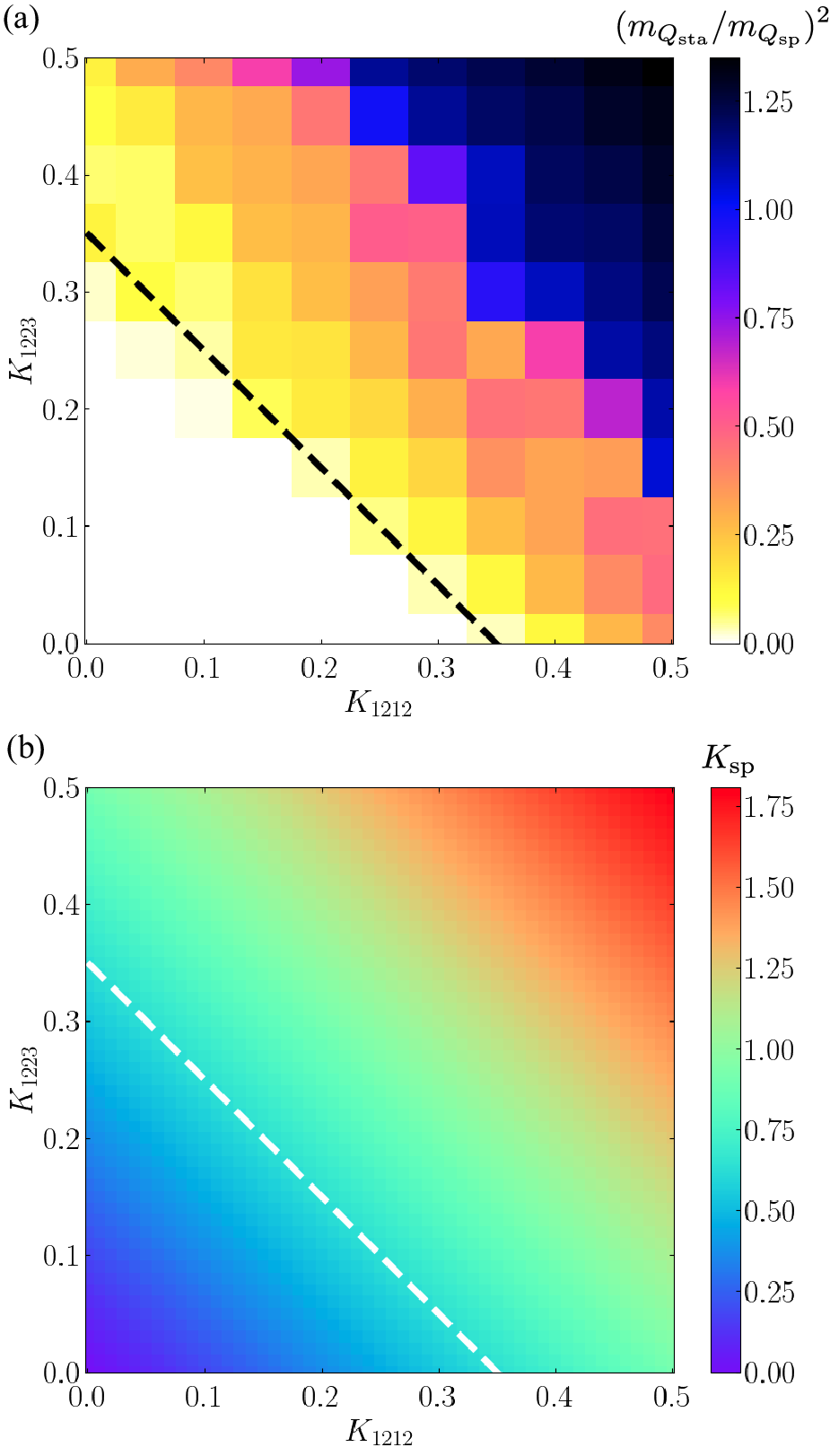} 
\caption{
\label{fig: realspace} 
(Color online) Contour plots of (a) $(m_{Q_{\rm sta}}/m_{Q_{\rm sp}})^2$ and (b) $K_{\rm sp}$ against the real-space four-spin interactions $K_{1212}$ and $K_{1223}$. 
The dashed lines denote the phase boundary between the single-$Q$ and double-$Q$ states. 
}
\end{center}
\end{figure}

In this section, we investigate the emergence of the double-$Q$ state based on the analysis of the real-space spin model. 
We consider the Hamiltonian given by 
\begin{align}
\label{eq: Hamreal}
\mathcal{H} = &\sum_n \sum_i  \sum_{\nu=x,y} \mathcal{J}_n S^\nu_i S^\nu_{i+n} \nonumber \\ 
&+ \sum_{i}
K_{1212} \left(\sum_{\nu=x,y} S^\nu_i S^\nu_{i+1}\right)^2 \nonumber \\
&+\sum_{i}K_{1223} \left(\sum_{\nu=x,y} S^\nu_i S^\nu_{i+1}\right)\left(\sum_{\nu=x,y}S^\nu_{i+1} S^\nu_{i+2}\right). 
\end{align}
The first term represents the bilinear exchange interaction, where $\mathcal{J}_n$ is parametrized as 
\begin{align}
\mathcal{J}_n = J_1 \frac{\cos (k_1n )}{k_1 n} + J_2 \frac{\cos (k_2n )}{k_2 n}, 
\end{align}
where the interaction includes further-neighbor contributions up to the 10th-neighbor spins ($n=10$). 
To consider the situation where the energies of the single-$Q$ spiral and staggered states are competitive, we set the interaction parameters as $J_1=-0.2247$, $J_2=-0.9876$, $k_1=0.1\pi$, and $k_2 = \pi$. 
This parameter set gives $J_{\rm sp} = 1$ and $J_{\rm sta}/J_{\rm sp}=0.4$ for $Q_{\rm sp}=0.1\pi$ in the momentum-resolved interaction~\cite{Okigami_PhysRevB.110.L220405}. 
The second and third terms represent the four-body interactions among two and three spins, respectively. 
We deal with the coupling constants $K_{1212}$ and $K_{1223}$ as phenomenological parameters. 

It should be noted that the real-space model in Eq.~(\ref{eq: Hamreal}) is not obtained through a direct Fourier transformation of the momentum-space Hamiltonian in Eq.~(\ref{eq: Ham}). 
Instead, Eq.~(\ref{eq: Hamreal}) is introduced as a phenomenological real-space effective model designed to capture the stability of the broken helix state. 
The long-range interaction $\mathcal{J}_n$ mimics an RKKY-type oscillatory exchange interaction consistent with two characteristic wave numbers $k_1$ and $k_2$, whereas the second- and third-line terms describe short-range four-body interactions that have no one-to-one correspondence to the biquadratic terms in Eq.~(\ref{eq: Ham}). 
This clarifies that the real-space model, constructed independently of the momentum-space model, is nevertheless capable of describing the same double-$Q$ broken helix state as the momentum-space formulation, as shown below.

We calculate the ground-state spin configuration by performing simulated annealing based on the Metropolis algorithm. 
The temperature scheduling is set as follows: 
The initial and final temperatures are $1$--$10$ and $0.01$, respectively. 
We gradually reduce the temperature via $10^6$--$10^7$ Monte Carlo steps.  
When the temperature reaches the final temperature, we further perform additional $10^6$ Monte Carlo sweeps for thermalization and measurement. 

Figure~\ref{fig: realspace}(a) shows the phase diagram when $K_{1212}$ and $K_{1223}$ are varied. 
The contour indicates $(m_{Q_{\rm sta}}/m_{Q_{\rm sp}})^2$, which distinguishes the single-$Q$ spiral state from the double-$Q$ state. 
It is noted that $\bm{Q}_{\rm sp}$ and $\bm{Q}_{\rm sta}$ slightly deviate from $k_1$, $k_2$ ($\lesssim 2\pi/L$) for some parameters due to the model's incommensurability, which might be attributed to the four-body interaction. 
In the region for small $K_{1212}$ and $K_{1223}$, the single-$Q$ spiral state is stabilized, whereas the double-$Q$ state appears by increasing $K_{1212}$ and $K_{1223}$. 
This tendency that large four-body interactions favor the double-$Q$ state is consistent with that in the momentum-based spin model in Fig.~\ref{fig: PD}. 
Indeed, as shown in Fig.~\ref{fig: realspace}(b), the biquadratic interaction $K_{\rm sp}$, which is evaluated from the Fourier transform of the second and third terms in Eq.~(\ref{eq: Hamreal}), becomes larger as $K_{1212}$ and $K_{1223}$ increase. 
Thus, the real-space spin model also supports that the biquadratic interaction in momentum space plays an important role in stabilizing the double-$Q$ state.

\section{Electronic band structure}
\label{sec: Electronic band structure}

In this section, we discuss the change of the electronic band structure under the single-$Q$ spiral and double-$Q$ orderings. 
To discuss a general tendency, we consider an itinerant electron model, where free electrons couple to localized spins via exchange coupling. 
The model Hamiltonian is given by 
\begin{align}
\label{eq: Ham_krep}
\mathcal{H}=&\sum_{k \sigma} \varepsilon_{k} c^{\dagger}_{k\sigma}c_{k\sigma} 
+ \frac{J_{\rm K}}{2} \sum_{k q\sigma\sigma'}
c^{\dagger}_{k\sigma}\bm{\sigma}_{\sigma \sigma'}c_{k+q\sigma'} \cdot \bm{S}_{q} \nonumber \\ 
&- \frac{H}{2} \sum_{k } 
(c^{\dagger}_{k\uparrow}c_{k\uparrow}-c^{\dagger}_{k\downarrow}c_{k\downarrow}) ,  
\end{align}
where $c^{\dagger}_{k\sigma}$ ($c_{k\sigma}$) is the creation (annihilation) operator of an itinerant electron at wave vector $k$ and spin $\sigma$, and $\bm{S}_{q}$ represents the Fourier transform of a localized spin $\bm{S}_i$, which is obtained by the simulated annealing in Sec.~\ref{sec: Multiple-$Q$ instability}. 
The first term represents the hopping term of itinerant electrons, where $\varepsilon_{k}$ is the energy dispersion represented as $\varepsilon_{k}=2t \cos k $ with the nearest-neighbor hopping $t=1$ as the energy unit, 
The second term stands for the exchange coupling between itinerant electron spins and localized spins with the magnitude of $J_{\rm K}$; $\bm{\sigma}=(\sigma^x,\sigma^y,\sigma^z)$ is the vector of Pauli matrices. 
The third term stands for the Zeeman coupling for itinerant electrons in the presence of an out-of-plane magnetic field with the magnitude of $H$. 
Although the spin configuration in $\bm{S}_q$ contanis no $z$-spin component, its noncollinear texture within the $xy$ plane nonetheless influences the $z$-spin polarization of the itinerant electrons, as discussed below.

\begin{figure}[tb!]
\begin{center}
\includegraphics[width=1.0\hsize]{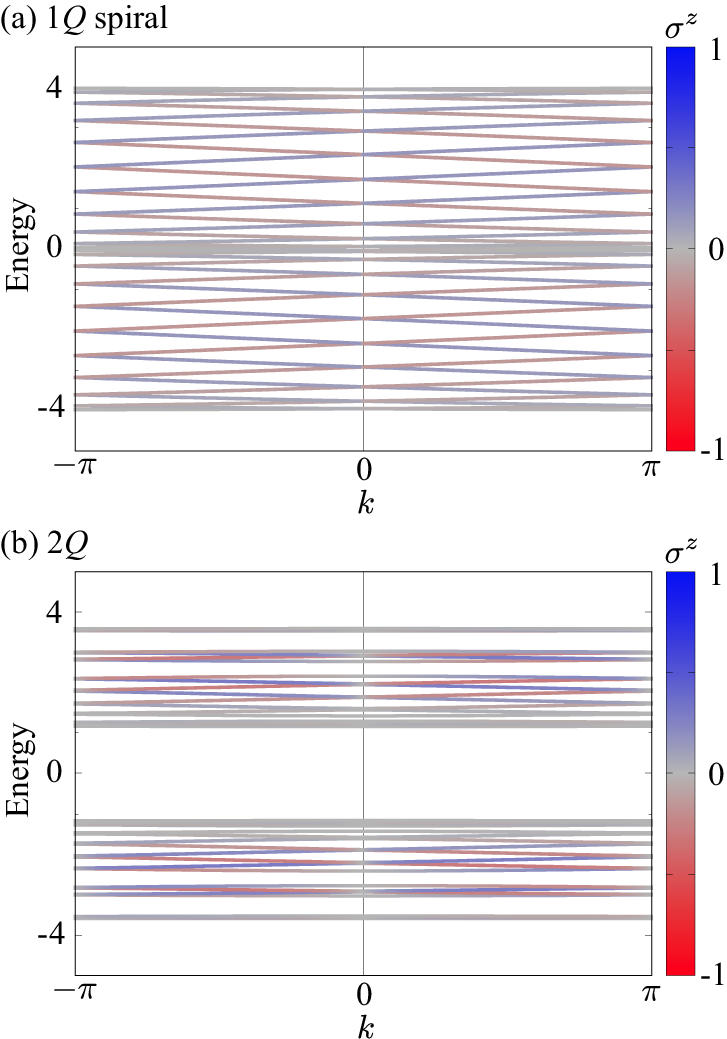} 
\caption{
\label{fig: band} 
(Color online) Electronic band structure in the absence of the magnetic field ($H=0$) under (a) the single-$Q$ spiral state at $Q_{\rm sp}=0.1\pi$ and (b) the double-$Q$ state at $Q_{\rm sp}=0.1\pi$, $K=0.6$, and $J_{\rm sta}=0.4$. 
The color represents the $z$-spin polarization. 
}
\end{center}
\end{figure}

\begin{figure}[tb!]
\begin{center}
\includegraphics[width=1.0\hsize]{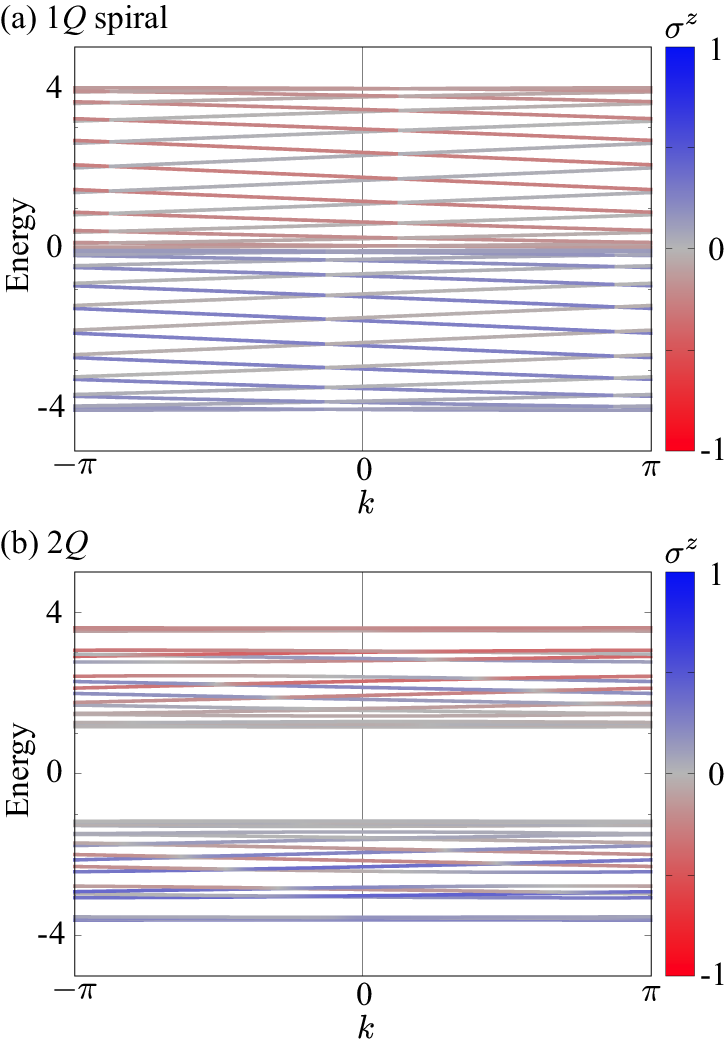} 
\caption{
\label{fig: band_Hz} 
(Color online) Electronic band structure in the presence of the magnetic field ($H=0.5$) under (a) the single-$Q$ spiral state at $Q_{\rm sp}=0.1\pi$ and (b) the double-$Q$ state at $Q_{\rm sp}=0.1\pi$, $K=0.6$, and $J_{\rm sta}=0.4$. 
The color represents the $z$-spin polarization. 
}
\end{center}
\end{figure}

Figure~\ref{fig: band}(a) shows the electronic band structure under the single-$Q$ spiral spin configuration with $Q_{\rm sp}=0.1\pi$ at zero magnetic field. 
Although the band shape seems to be symmetric with respect to $k=0$, the out-of-plane spin component is polarized in an antisymmetric way. 
In other words, the antisymmetric spin splitting in the form of $k \sigma^z$ occurs in the single-$Q$ state, where $\sigma^z$ at each wave number and each band is calculated as $\bra{nk}\sigma^z \ket{nk}$ by using the eigenstate $\ket{nk}$ with the band index $n$ and wave vector $k$. 
It is noted that this antisymmetric spin splitting is purely driven by the magnetic order, since the present model does not include the effect of the relativistic spin--orbit coupling~\cite{Hayami_PhysRevB.101.220403, Brekke_PhysRevLett.133.236703, Sukhachov_PhysRevB.110.205114, Soori_PhysRevB.111.165413, Nagae_PhysRevB.111.174519, Hodt_PhysRevB.111.205416}. 
From the microscopic viewpoint, the antisymmetric spin splitting is caused by nonzero vector spin chirality~\cite{Hayami_PhysRevB.105.024413}.
This is because the vector spin chirality $(\bm{S}_i \times \bm{S}_j)$ is, from symmetry considerations, associated with the product of the three-dimensional wave vector and spin $k_{\mu}\sigma^\nu$ depending on the components $\mu,\nu=x,y,z$, particularly in general three-dimensional systems. 
In such a situation, the $z$ component of $(\bm{S}_i \times \bm{S}_j)$ in the one-dimensional chain corresponds to the $k_x \sigma_z - k_z \sigma_x$ at the symmetry level, where both quantities are characterized by the $y$-directional polar field breaking the spatial inversion symmetry like the electric polarization~\cite{Katsura_PhysRevLett.95.057205, Mostovoy_PhysRevLett.96.067601, SergienkoPhysRevB.73.094434, Harris_PhysRevB.73.184433,tokura2014multiferroics}. 
Similar antisymmetric spin splitting is also found in the double-$Q$ state, as its spin configuration includes the single-$Q$ spiral one, as shown in Fig.~\ref{fig: band}(b). 
In addition, the gap-opening tendency at half filling appears owing to the staggered component. 

When the magnetic field $H$ is turned on, additional band modulations occur in both single-$Q$ spiral and double-$Q$ states; the asymmetric band modulation in terms of $k=0$ occurs, as shown in Figs.~\ref{fig: band_Hz}(a) and \ref{fig: band_Hz}(b). 
This asymmetric band modulation is caused by the noncoplanar spin configuration, where the spatial alignment of the local scalar spin chirality breaks both the spatial inversion and time-reversal symmetries~\cite{Hayami_PhysRevB.106.014420, Hayami_doi:10.7566/JPSJ.91.094704}. 
From the symmetry viewpoint, the appearance of the asymmetric band modulation is natural, since the simultaneous existence of the $y$-directional polar field and the $z$-directional magnetic field gives rise to the magnetic toroidal moment along the $x$ direction~\cite{dubovik1975multipole,dubovik1990toroid,gorbatsevich1994toroidal, Spaldin_0953-8984-20-43-434203, kopaev2009toroidal}, which leads to $x$-directional asymmetric band deformations in electron~\cite{Yanase_JPSJ.83.014703, Hayami_doi:10.7566/JPSJ.84.064717} and magnon~\cite{Miyahara_JPSJ.81.023712, Miyahara_PhysRevB.89.195145, Hayami_doi:10.7566/JPSJ.85.053705, Matsumoto_PhysRevB.101.224419, Aoyama_PhysRevB.111.144413} systems. 
Thus, the second-order nonlinear nonreciprocal conductivity defined as $J_x = \sigma_{xxx} E_x^2$ is expected in both orderings without the relativistic spin--orbit coupling~\cite{Yatsushiro_PhysRevB.105.155157, Hayami_PhysRevB.106.014420}.

\section{Summary}
\label{sec:summary}

To summarize, we theoretically explored a new class of multiple-$Q$ spin textures that emerge in one-dimensional systems through the interference between spiral and staggered magnetic modulations with symmetry-unrelated ordering wave vectors. 
Motivated by recent experimental observations of broken helix states in materials such as EuIn$_2$As$_2$, we examined the stability of the double-$Q$ state; our model does not aim to reproduce the microscopic interactions of EuIn$_2$As$_2$, but rather to identify the general mechanisms that can stabilize such a state.
For that purpose, we constructed a momentum-resolved spin model including bilinear and biquadratic exchange interactions and analyzed its ground state by performing simulated annealing. 
As a result, we identified the parameter regimes where the double-$Q$ state becomes energetically favorable. 
We showed that the positive biquadratic interaction and the competition between finite-$Q$ spiral and staggered exchange interactions play a key role in stabilizing the double-$Q$ state. 
We also confirmed the importance of a positive biquadratic interaction by analyzing the real-space spin model. 
The resulting spin configuration leads to antisymmetric spin-split electronic bands even in the absence of spin--orbit coupling, and exhibits asymmetric band modulation under an applied magnetic field. 
Our findings provide insight into the stabilization mechanism of symmetry-unrelated multiple-$Q$ states in centrosymmetric magnets, which stimulates further experimental findings.

\begin{acknowledgments}
We acknowledge M. Gen for enlightening discussions in the early stage of this
study. 
This research was supported by JSPS KAKENHI Grants Numbers JP22H00101, JP22H01183, JP23H04869, JP23K03288, and by JST CREST (JPMJCR23O4) and JST FOREST (JPMJFR2366). 
K. O. was supported by JST SPRING, Grant Number JPMJSP2108. 
Parts of the numerical calculations were performed in the supercomputing systems in ISSP, the University of Tokyo.
\end{acknowledgments}

\appendix

\bibliographystyle{jpsj}

\begin{thebibliography}{10}

\bibitem{Bak_PhysRevLett.40.800}
P.~Bak and B.~Lebech, Phys. Rev. Lett. {\bfseries 40},  800 (1978).

\bibitem{McEwen_PhysRevB.34.1781}
K.~A. McEwen and M.~B. Walker, Phys. Rev. B {\bfseries 34},  1781 (1986).

\bibitem{zochowski1986thermal}
S.~Zochowski and K.~McEwen, J. Magn. Magn. Mater. {\bfseries 54},  515 (1986).

\bibitem{forgan1990magnetic}
E.~Forgan, B.~Rainford, S.~Lee, J.~Abell, and Y.~Bi, J. Phys.: Condens. Matter
  {\bfseries 2},  10211 (1990).

\bibitem{Longfield_PhysRevB.66.054417}
M.~J. Longfield, J.~A. Paix\~ao, N.~Bernhoeft, and G.~H. Lander, Phys. Rev. B
  {\bfseries 66},  054417 (2002).

\bibitem{Bernhoeft_PhysRevB.69.174415}
N.~Bernhoeft, J.~A. Paix\~ao, C.~Detlefs, S.~B. Wilkins, P.~Javorsk\'y,
  E.~Blackburn, and G.~H. Lander, Phys. Rev. B {\bfseries 69},  174415 (2004).

\bibitem{stewart2004phase}
J.~Stewart, G.~Ehlers, A.~Wills, S.~T. Bramwell, and J.~Gardner, J. Phys.:
  Condens. Matter {\bfseries 16},  L321 (2004).

\bibitem{Watson_PhysRevB.53.726}
D.~Watson, E.~M. Forgan, W.~J. Nuttall, W.~G. Stirling, and D.~Fort, Phys. Rev.
  B {\bfseries 53},  726 (1996).

\bibitem{Harris_PhysRevB.74.134411}
A.~B. Harris and J.~Schweizer, Phys. Rev. B {\bfseries 74},  134411 (2006).

\bibitem{nagaosa2013topological}
N.~Nagaosa and Y.~Tokura, Nat. Nanotechnol. {\bfseries 8},  899 (2013).

\bibitem{Ohgushi_PhysRevB.62.R6065}
K.~Ohgushi, S.~Murakami, and N.~Nagaosa, Phys. Rev. B {\bfseries 62},  R6065
  (2000).

\bibitem{taguchi2001spin}
Y.~Taguchi, Y.~Oohara, H.~Yoshizawa, N.~Nagaosa, and Y.~Tokura, Science
  {\bfseries 291},  2573 (2001).

\bibitem{tatara2002chirality}
G.~Tatara and H.~Kawamura, J. Phys. Soc. Jpn. {\bfseries 71},  2613 (2002).

\bibitem{Neubauer_PhysRevLett.102.186602}
A.~Neubauer, C.~Pfleiderer, B.~Binz, A.~Rosch, R.~Ritz, P.~G. Niklowitz, and
  P.~B\"oni, Phys. Rev. Lett. {\bfseries 102},  186602 (2009).

\bibitem{Hamamoto_PhysRevB.92.115417}
K.~Hamamoto, M.~Ezawa, and N.~Nagaosa, Phys. Rev. B {\bfseries 92},  115417
  (2015).

\bibitem{nakazawa2018topological}
K.~Nakazawa, M.~Bibes, and H.~Kohno, J. Phys. Soc. Jpn. {\bfseries 87},  033705
  (2018).

\bibitem{tai2022distinguishing}
L.~Tai, B.~Dai, J.~Li, H.~Huang, S.~K. Chong, K.~L. Wong, H.~Zhang, P.~Zhang,
  P.~Deng, C.~Eckberg, G.~Qiu, H.~He, D.~Wu, S.~Xu, A.~Davydov, R.~Wu, and
  K.~L. Wang, ACS nano {\bfseries 16},  17336 (2022).

\bibitem{Katsura_PhysRevLett.95.057205}
H.~Katsura, N.~Nagaosa, and A.~V. Balatsky, Phys. Rev. Lett. {\bfseries 95},
  057205 (2005).

\bibitem{Mostovoy_PhysRevLett.96.067601}
M.~Mostovoy, Phys. Rev. Lett. {\bfseries 96},  067601 (2006).

\bibitem{SergienkoPhysRevB.73.094434}
I.~A. Sergienko and E.~Dagotto, Phys. Rev. B {\bfseries 73},  094434 (2006).

\bibitem{Harris_PhysRevB.73.184433}
A.~B. Harris, T.~Yildirim, A.~Aharony, and O.~Entin-Wohlman, Phys. Rev. B
  {\bfseries 73},  184433 (2006).

\bibitem{tokura2014multiferroics}
Y.~Tokura, S.~Seki, and N.~Nagaosa, Rep. Prog. Phys. {\bfseries 77},  076501
  (2014).

\bibitem{cardias2020first}
R.~Cardias, A.~Szilva, M.~Bezerra-Neto, M.~Ribeiro, A.~Bergman, Y.~O. Kvashnin,
  J.~Fransson, A.~Klautau, O.~Eriksson, and L.~Nordstr{\"o}m, Sci. Rep.
  {\bfseries 10},  20339 (2020).

\bibitem{Hayami_PhysRevB.105.104428}
S.~Hayami and R.~Yambe, Phys. Rev. B {\bfseries 105},  104428 (2022).

\bibitem{Bhowal_PhysRevLett.128.227204}
S.~Bhowal and N.~A. Spaldin, Phys. Rev. Lett. {\bfseries 128},  227204 (2022).

\bibitem{Shapiro_PhysRevLett.43.1748}
S.~M. Shapiro, E.~Gurewitz, R.~D. Parks, and L.~C. Kupferberg, Phys. Rev. Lett.
  {\bfseries 43},  1748 (1979).

\bibitem{bak1980theory}
P.~Bak and M.~H. Jensen, J. Phys. C: Solid State Phys. {\bfseries 13},  L881
  (1980).

\bibitem{Forgan_PhysRevLett.62.470}
E.~M. Forgan, E.~P. Gibbons, K.~A. McEwen, and D.~Fort, Phys. Rev. Lett.
  {\bfseries 62},  470 (1989).

\bibitem{hayami2024stabilization}
S.~Hayami and R.~Yambe, Mater. Today Quantum {\bfseries 3},  100010 (2024).

\bibitem{Muhlbauer_2009skyrmion}
S.~M{\"u}hlbauer, B.~Binz, F.~Jonietz, C.~Pfleiderer, A.~Rosch, A.~Neubauer,
  R.~Georgii, and P.~B{\"o}ni, Science {\bfseries 323},  915 (2009).

\bibitem{yu2010real}
X.~Z. Yu, Y.~Onose, N.~Kanazawa, J.~H. Park, J.~H. Han, Y.~Matsui, N.~Nagaosa,
  and Y.~Tokura, Nature {\bfseries 465},  901 (2010).

\bibitem{rossler2006spontaneous}
U.~K. R{\"o}{\ss}ler, A.~N. Bogdanov, and C.~Pfleiderer, Nature {\bfseries
  442},  797 (2006).

\bibitem{Binz_PhysRevLett.96.207202}
B.~Binz, A.~Vishwanath, and V.~Aji, Phys. Rev. Lett. {\bfseries 96},  207202
  (2006).

\bibitem{Binz_PhysRevB.74.214408}
B.~Binz and A.~Vishwanath, Phys. Rev. B {\bfseries 74},  214408 (2006).

\bibitem{Yi_PhysRevB.80.054416}
S.~D. Yi, S.~Onoda, N.~Nagaosa, and J.~H. Han, Phys. Rev. B {\bfseries 80},
  054416 (2009).

\bibitem{khanh2020nanometric}
N.~D. Khanh, T.~Nakajima, X.~Yu, S.~Gao, K.~Shibata, M.~Hirschberger,
  Y.~Yamasaki, H.~Sagayama, H.~Nakao, L.~Peng, K.~Nakajima, R.~Takagi, T.-h.
  Arima, Y.~Tokura, and S.~Seki, Nat. Nanotechnol. {\bfseries 15},  444 (2020).

\bibitem{singh2023transition}
D.~Singh, Y.~Fujishiro, S.~Hayami, S.~H. Moody, T.~Nomoto, P.~R. Baral,
  V.~Ukleev, R.~Cubitt, N.-J. Steinke, D.~J. Gawryluk, E.~Pomjakushina,
  Y.~{\=O}nuki, R.~Arita, Y.~Tokura, N.~Kanazawa, and J.~S. White, Nat. Commun.
  {\bfseries 14},  8050 (2023).

\bibitem{riberolles2021magnetic}
S.~X. Riberolles, T.~V. Trevisan, B.~Kuthanazhi, T.~Heitmann, F.~Ye,
  D.~Johnston, S.~Bud’ko, D.~Ryan, P.~Canfield, A.~Kreyssig, A.~Vishwanath,
  R.~J. McQueeney, L.~L. Wang, P.~P. Orth, and B.~G. Ueland, Nat. Commun.
  {\bfseries 12},  999 (2021).

\bibitem{soh2023understanding}
J.-R. Soh, A.~Bombardi, F.~Mila, M.~C. Rahn, D.~Prabhakaran, S.~Francoual,
  H.~M. R{\o}nnow, and A.~T. Boothroyd, Nat. Commun. {\bfseries 14},  3387
  (2023).

\bibitem{Donoway_PhysRevX.14.031013}
E.~Donoway, T.~V. Trevisan, A.~Liebman-Pel\'aez, R.~P. Day, K.~Yamakawa,
  Y.~Sun, J.~R. Soh, D.~Prabhakaran, A.~T. Boothroyd, R.~M. Fernandes, J.~G.
  Analytis, J.~E. Moore, J.~Orenstein, and V.~Sunko, Phys. Rev. X {\bfseries
  14},  031013 (2024).

\bibitem{takeda2024incommensurate}
H.~Takeda, J.~Yan, Z.~Jiang, X.~Luo, Y.~Sun, and M.~Yamashita, npj Quantum
  Materials {\bfseries 9},  67 (2024).

\bibitem{Gen_PhysRevB.111.L081109}
M.~Gen, Y.~Fujishiro, K.~Okigami, S.~Hayami, M.~T. Birch, K.~Adachi,
  D.~Hashizume, T.~Kurumaji, H.~Sagayama, H.~Nakao, Y.~Tokura, and T.-h. Arima,
  Phys. Rev. B {\bfseries 111},  L081109 (2025).

\bibitem{dzyaloshinsky1958thermodynamic}
I.~Dzyaloshinsky, J. Phys. Chem. Solids {\bfseries 4},  241 (1958).

\bibitem{moriya1960anisotropic}
T.~Moriya, Phys. Rev. {\bfseries 120},  91 (1960).

\bibitem{Hayami_PhysRevB.95.224424}
S.~Hayami, R.~Ozawa, and Y.~Motome, Phys. Rev. B {\bfseries 95},  224424
  (2017).

\bibitem{Yambe_PhysRevB.106.174437}
R.~Yambe and S.~Hayami, Phys. Rev. B {\bfseries 106},  174437 (2022).

\bibitem{Hayami_PhysRevB.99.094420}
S.~Hayami and Y.~Motome, Phys. Rev. B {\bfseries 99},  094420 (2019).

\bibitem{hayami2020multiple}
S.~Hayami, J. Magn. Magn. Mater. {\bfseries 513},  167181 (2020).

\bibitem{hayami2021phase}
S.~Hayami, T.~Okubo, and Y.~Motome, Nat. Commun. {\bfseries 12},  6927 (2021).

\bibitem{hayami2022multiple}
S.~Hayami, J. Phys. Soc. Jpn. {\bfseries 91},  023705 (2022).

\bibitem{Hayami_PhysRevB.104.094425}
S.~Hayami and R.~Yambe, Phys. Rev. B {\bfseries 104},  094425 (2021).

\bibitem{Okumura_PhysRevB.101.144416}
S.~Okumura, S.~Hayami, Y.~Kato, and Y.~Motome, Phys. Rev. B {\bfseries 101},
  144416 (2020).

\bibitem{Okumura_doi:10.7566/JPSJ.91.093702}
S.~Okumura, S.~Hayami, Y.~Kato, and Y.~Motome, J. Phys. Soc. Jpn. {\bfseries
  91},  093702 (2022).

\bibitem{Ruderman}
M.~A. Ruderman and C.~Kittel, Phys. Rev. {\bfseries 96},  99 (1954).

\bibitem{Kasuya}
T.~Kasuya, Prog. Theor. Phys. {\bfseries 16},  45 (1956).

\bibitem{Yosida1957}
K.~Yosida, Phys. Rev. {\bfseries 106},  893 (1957).

\bibitem{Akagi_PhysRevLett.108.096401}
Y.~Akagi, M.~Udagawa, and Y.~Motome, Phys. Rev. Lett. {\bfseries 108},  096401
  (2012).

\bibitem{Hayami_PhysRevB.90.060402}
S.~Hayami and Y.~Motome, Phys. Rev. B {\bfseries 90},  060402(R) (2014).

\bibitem{takagi2018multiple}
R.~Takagi, J.~White, S.~Hayami, R.~Arita, D.~Honecker, H.~R{\o}nnow, Y.~Tokura,
  and S.~Seki, Sci. Adv. {\bfseries 4},  eaau3402 (2018).

\bibitem{kakihana2018giant}
M.~Kakihana, D.~Aoki, A.~Nakamura, F.~Honda, M.~Nakashima, Y.~Amako,
  S.~Nakamura, T.~Sakakibara, M.~Hedo, T.~Nakama, and Y.~Onuki, J. Phys. Soc.
  Jpn. {\bfseries 87},  023701 (2018).

\bibitem{kaneko2019unique}
K.~Kaneko, M.~D. Frontzek, M.~Matsuda, A.~Nakao, K.~Munakata, T.~Ohhara,
  M.~Kakihana, Y.~Haga, M.~Hedo, T.~Nakama, and Y.~Onuki, J. Phys. Soc. Jpn.
  {\bfseries 88},  013702 (2019).

\bibitem{tabata2019magnetic}
C.~Tabata, T.~Matsumura, H.~Nakao, S.~Michimura, M.~Kakihana, T.~Inami,
  K.~Kaneko, M.~Hedo, T.~Nakama, and Y.~{\=O}nuki, J. Phys. Soc. Jpn.
  {\bfseries 88},  093704 (2019).

\bibitem{kakihana2019unique}
M.~Kakihana, D.~Aoki, A.~Nakamura, F.~Honda, M.~Nakashima, Y.~Amako,
  T.~Takeuchi, H.~Harima, M.~Hedo, T.~Nakama, and Y.~Onuki, J. Phys. Soc. Jpn.
  {\bfseries 88},  094705 (2019).

\bibitem{hayami2021field}
S.~Hayami and R.~Yambe, J. Phys. Soc. Jpn. {\bfseries 90},  073705 (2021).

\bibitem{Hayami_PhysRevB.103.054422}
S.~Hayami and Y.~Motome, Phys. Rev. B {\bfseries 103},  054422 (2021).

\bibitem{amoroso2020spontaneous}
D.~Amoroso, P.~Barone, and S.~Picozzi, Nat. Commun. {\bfseries 11},  5784
  (2020).

\bibitem{yambe2021skyrmion}
R.~Yambe and S.~Hayami, Sci. Rep. {\bfseries 11},  11184 (2021).

\bibitem{Wang_PhysRevB.103.104408}
Z.~Wang, Y.~Su, S.-Z. Lin, and C.~D. Batista, Phys. Rev. B {\bfseries 103},
  104408 (2021).

\bibitem{Utesov_PhysRevB.103.064414}
O.~I. Utesov, Phys. Rev. B {\bfseries 103},  064414 (2021).

\bibitem{amoroso2021tuning}
D.~Amoroso, P.~Barone, and S.~Picozzi, Nanomaterials {\bfseries 11},  1873
  (2021).

\bibitem{Utesov_PhysRevB.105.054435}
O.~I. Utesov, Phys. Rev. B {\bfseries 105},  054435 (2022).

\bibitem{Luttinger_PhysRev.70.954}
J.~M. Luttinger and L.~Tisza, Phys. Rev. {\bfseries 70},  954 (1946).

\bibitem{Ozawa_doi:10.7566/JPSJ.85.103703}
R.~Ozawa, S.~Hayami, K.~Barros, G.-W. Chern, Y.~Motome, and C.~D. Batista, J.
  Phys. Soc. Jpn. {\bfseries 85},  103703 (2016).

\bibitem{Okigami_PhysRevB.110.L220405}
K.~Okigami and S.~Hayami, Phys. Rev. B {\bfseries 110},  L220405 (2024).

\bibitem{Hayami_PhysRevB.101.220403}
S.~Hayami, Y.~Yanagi, and H.~Kusunose, Phys. Rev. B {\bfseries 101},  220403(R)
  (2020).

\bibitem{Brekke_PhysRevLett.133.236703}
B.~Brekke, P.~Sukhachov, H.~G. Giil, A.~Brataas, and J.~Linder, Phys. Rev.
  Lett. {\bfseries 133},  236703 (2024).

\bibitem{Sukhachov_PhysRevB.110.205114}
P.~Sukhachov and J.~Linder, Phys. Rev. B {\bfseries 110},  205114 (2024).

\bibitem{Soori_PhysRevB.111.165413}
A.~Soori, Phys. Rev. B {\bfseries 111},  165413 (2025).

\bibitem{Nagae_PhysRevB.111.174519}
Y.~Nagae, L.~Katayama, and S.~Ikegaya, Phys. Rev. B {\bfseries 111},  174519
  (2025).

\bibitem{Hodt_PhysRevB.111.205416}
E.~W. Hodt, H.~Bentmann, and J.~Linder, Phys. Rev. B {\bfseries 111},  205416
  (2025).

\bibitem{Hayami_PhysRevB.105.024413}
S.~Hayami, Phys. Rev. B {\bfseries 105},  024413 (2022).

\bibitem{Hayami_PhysRevB.106.014420}
S.~Hayami and M.~Yatsushiro, Phys. Rev. B {\bfseries 106},  014420 (2022).

\bibitem{Hayami_doi:10.7566/JPSJ.91.094704}
S.~Hayami and M.~Yatsushiro, J. Phys. Soc. Jpn. {\bfseries 91},  094704 (2022).

\bibitem{dubovik1975multipole}
V.~Dubovik and A.~Cheshkov, Sov. J. Part. Nucl {\bfseries 5},  318 (1975).

\bibitem{dubovik1990toroid}
V.~Dubovik and V.~Tugushev, Phys. Rep. {\bfseries 187},  145 (1990).

\bibitem{gorbatsevich1994toroidal}
A.~Gorbatsevich and Y.~V. Kopaev, Ferroelectrics {\bfseries 161},  321 (1994).

\bibitem{Spaldin_0953-8984-20-43-434203}
N.~A. Spaldin, M.~Fiebig, and M.~Mostovoy, J. Phys.: Condens. Matter {\bfseries
  20},  434203 (2008).

\bibitem{kopaev2009toroidal}
Y.~V. Kopaev, Physics-Uspekhi {\bfseries 52},  1111 (2009).

\bibitem{Yanase_JPSJ.83.014703}
Y.~Yanase, J. Phys. Soc. Jpn. {\bfseries 83},  014703 (2014).

\bibitem{Hayami_doi:10.7566/JPSJ.84.064717}
S.~Hayami, H.~Kusunose, and Y.~Motome, J. Phys. Soc. Jpn. {\bfseries 84},
  064717 (2015).

\bibitem{Miyahara_JPSJ.81.023712}
S.~Miyahara and N.~Furukawa, J. Phys. Soc. Jpn. {\bfseries 81},  023712 (2012).

\bibitem{Miyahara_PhysRevB.89.195145}
S.~Miyahara and N.~Furukawa, Phys. Rev. B {\bfseries 89},  195145 (2014).

\bibitem{Hayami_doi:10.7566/JPSJ.85.053705}
S.~Hayami, H.~Kusunose, and Y.~Motome, J. Phys. Soc. Jpn. {\bfseries 85},
  053705 (2016).

\bibitem{Matsumoto_PhysRevB.101.224419}
T.~Matsumoto and S.~Hayami, Phys. Rev. B {\bfseries 101},  224419 (2020).

\bibitem{Aoyama_PhysRevB.111.144413}
K.~Aoyama and H.~Kawamura, Phys. Rev. B {\bfseries 111},  144413 (2025).

\bibitem{Yatsushiro_PhysRevB.105.155157}
M.~Yatsushiro, R.~Oiwa, H.~Kusunose, and S.~Hayami, Phys. Rev. B {\bfseries
  105},  155157 (2022).

\end{thebibliography}

\end{document}